\documentclass[a4paper,english,conference]{IEEEtran}
\usepackage[T1]{fontenc}
\usepackage[latin9]{inputenc}

\usepackage{epsfig}
\usepackage{graphicx}
\usepackage{comment}
\usepackage[colorinlistoftodos]{todonotes}
\usepackage{algorithm}
\usepackage[noend]{algpseudocode}
\usepackage{amssymb}
\usepackage{amsmath}
\usepackage[labelfont=small,textfont=small]{caption}
\usepackage{xcolor}
\usepackage{balance}

\DeclareRobustCommand*{\IEEEauthorrefmark}[1]{%
	\raisebox{0pt}[0pt][0pt]{\textsuperscript{\footnotesize #1}}%
}

\begin{document}
\author{}
\title{}
\title{Transforming Collaboration Data into Network Layers for Enhanced Analytics}

\author{\IEEEauthorblockN{Saharnaz E. Dilmaghani\IEEEauthorrefmark{1}, Apivadee Piyatumrong\IEEEauthorrefmark{2}, Pascal Bouvry\IEEEauthorrefmark{1}, Matthias R. Brust\IEEEauthorrefmark{1} }\IEEEauthorblockA{\IEEEauthorrefmark{1}Interdisciplinary Centre for Security, Reliability and Trust (SnT), University of Luxembourg, Luxembourg\\
		\IEEEauthorrefmark{2}National Electronics and Computer Technology Center (NECTEC), Thailand \\
		saharnaz.dilmaghani@uni.lu, apivadee.piy@nectec.or.th, pascal.bouvry@uni.lu, matthias.brust@uni.lu}}

\maketitle

\begin{abstract}
We consider the problem of automatically generating networks from data of collaborating researchers. The objective is to apply network analysis on the resulting network layers to reveal supplemental patterns and insights of the research collaborations. In this paper, we describe our data-to-networks method, which automatically generates a set of logical network layers from the relational input data using a linkage threshold. We, then, use a series of network metrics to analyze the impact of the linkage threshold on the individual network layers. Moreover, results from the network analysis also provide beneficial information to improve the network visualization. We demonstrate the feasibility and impact of our approach using real-world collaboration data. We discuss how the produced network layers can reveal insights and patterns to direct the data analytics more intelligently.


\end{abstract}

\section{Introduction}\label{sec:intro}

Network structures have drawn significant attention in big data due to the possibility to apply network theory and analysis to obtain extra insights from the data. Networks are ubiquitous \cite{shirinivas2010applications} in research areas from biology and neuroscience (e.g., brain networks \cite{bassett2018nature}) to modeling and analyzing galaxy distributions \cite{hong2016discriminating}, and quantifying reputation in art \cite{fraiberger2018quantifying}. These examples are use cases where network layers play an important role to represent and analyze the data. 

Networks offer several advantages. They provide adaptability for dynamic structures benefiting from local principles on nodes \cite{brust2007adaptive}. Additionally, network analysis discloses valuable information for data visualization by defining an appropriate link definition. Networks often provide computationally efficient algorithms with lower complexity in comparison to a tabular structure \cite{davisbig}. Furthermore, data transformed into network structures can help providing evidence for missing information \cite{yang2016predicting,pan2016predicting} as well as predicting forthcoming events \cite{sha2018network}. Besides, there are numerous algorithms that can be applied on networks such as \textit{Louvain}'s algorithm which is a community detection algorithm \cite{blondel2008fast} and \textit{Page Rank}, that identifies the most influential object within a network.

With these advantages of networks, we are confronted with the challenge on how to transform relational data into appropriate networks, which exhibit advantages for data analytics as well as data visualization. The challenge is twofold: It is not only on how to represent the elements of a network, but also the specific construction principles, since, for each dataset, there are numerous ways be transformed into a network representation \cite{butts2009revisiting}. Each network reveals a particular perspective on the dataset. In this study, we investigate different linkage thresholds for the transformation of a collaboration dataset into networks.

In this paper, we propose a method that transforms collaboration data to network layers. Our approach favors scientific projects as nodes with links generated by using a specific linkage threshold. We apply our method on real-world data that describes collaboration of a renowned research institute. Our study uses different metrics to determine the influence of the network structure on the network properties. Additionally, we obtain results that provide information on better visualizing the produced networks and a possible interface to include privacy mechanisms into data analytics \cite{wu2016privacy}.
The remainder of this paper is structured as follows. Section~\ref{sec:related} investigates related work. We describe our method to complement data with network layers in Section~\ref{sec:method}. The experiment setup designed to analyze the proposed algorithm is represented in Section~\ref{sec:setup}. Section~\ref{sec:results} discusses the outcome and the opportunities for future work. Finally, Section~\ref{sec:conclusion} concludes the paper.

\section{Related Work}\label{sec:related}

The challenge of converting data to networks is a well-known issue when it comes to geographical data \cite{sapatialdata1,sapatialdata2}. Graph studies on spatial data reveal valuable information on route networks, complex urban systems \cite{peiravian2014spatial} and the relationship between different urban areas \cite{zhong2014detecting}. Nevertheless, the raw dataset on geographical information is not enough by its own to conduct proper graph studies. Karduni et al. \cite{karduni2016protocol} focuses on this challenge and has introduced an approach for geographical systems by defining a protocol describing the network properties to convert the spatial polyline data into a network.

In general, some of the studies have stressed the inference of links from relational data to design a network out of the relational data. Casiraghi et al. \cite{casiraghi2017relational} developed a generalized hypergeometric ensembles approach to address the problem of inferring connections within relational data. The study represents a perspective of link prediction while applying predictive analysis. From a similar point of view, Xiang et al. \cite{xiang2010modeling} established a link-based latent variable model to infer the friendship relations within a social interaction. In addition, in another study \cite{schein2015bayesian} the international relations from a dataset consists of the news from different countries are extracted by a tensor factorization technique. Moreover, Akbas et al. \cite{akbas2013social} proposed a social network generation by proposing a model based on various interactions (e.g., phone calls) considering smartphone data. Given a weight to each type of interaction, the authors define a link value as the combination of various interaction types. Akbas et al. followed up their study on network generation from interaction patterns by studies on how to infer social networks of animal groups \cite{brust2011a,brust2011b,AKBAS2015207}. Initiating from these studies, in particular from \cite{akbas2013social}, we followed a similar approach in order to define the linkage threshold to generate the network layers \cite{andronache2007hycast, chen2013modeling}.

Considering the collaboration data, Newman \cite{newman2001scientific} has established networks considering authors and their collaboration on scientific papers. The scientific collaboration networks have been also studied in a particular network structure, \textit{hypergraphs}, by Ouvrard et. al. \cite{ouvrard2017networks}. The authors emphasized on enhancing the visualization of these networks considering network properties.

\section{Proposed Methodology: Data-to-Network Layers}\label{sec:method}
We establish a method with the purpose to convert the relational data of research collaborations into network layers by describing a set of nodes and a linkage threshold to define the connections between nodes. Considering the thresholds we then generate different network layers and investigate each network with the network measurements. Using outcomes of the last phase we then visualize the networks for a better understanding of the relationship in our data. 

For this study, we have access to a real-world dataset of collaborative projects within National Electronics and Computer Technology Center (NECTEC). The dataset consists of various projects that have been completed with certain members of the institute. Investigating on the relationships and extracting the collaboration patterns are the outcomes that the networks will provide in order to improve performance of the research projects by using the possible resources inside the organization. We, therefore, retrieved different features from the dataset to feed our method and generate the network layers.

In this study, our focus is particularly on the projects of the institute. Thus, we consider the elements of the network such that the projects are nodes of the network and the common members between the projects are represented as the links of the network. Each project is identified with an $id$, and a team of $m$ members who contributed to the project, whereas the project members can participate in different projects at the same time. Contribution percentage is a feature extracted from the dataset that quantifies the contribution of a member within a project.

We calculate the linkage threshold by looking at certain features from the dataset: (1) the project members, and (2) the contribution percentage of each project member in a project. Let $m^i_k$, $c^i_k$ be the $k$-th member of project $i$ and the contribution percentage of the member in the project, respectively. Let $m^{i} = (m^i_1, m^i_2, \ldots)$ and $c^{i} = (c^i_1, c^i_2, \ldots) $ of project $i \in \mathcal{ID}$. Next, we define $\mathcal{P}$ as the combination of features chosen to explain the explain linkage threshold, i.e., $\mathcal{P} = \bigcup_i(m^i, c^i)$. Then, the common members between two projects (e.g., $i$, $j$) is denoted as $\mathcal{M}_{ij} = \{ (i,j) \in \mathcal{ID} \, | \, m^{i} \cap m^{j}\}$. Finally, considering the aforementioned parameters, the linkage threshold is defined as

\begin{gather}\label{eq:linkagethreshold}
 \mathcal{T}_{ij} = \frac{1}{n_{ij}} \sum_{\forall k \in \mathcal{M}_{ij}} \dfrac{c^i_k + c^j_k}{2}, \nonumber\\[1ex]
 \mathcal{T} = \bigcup_{i,j \in \mathcal{ID}}(\mathcal{T}_{ij}),
\end{gather}

where $\mathcal{T}_{ij}$ is the linkage threshold within the projects $(i,j)$, $n_{ij}$ is the number of common members calculated in $\mathcal{M}_{ij}$, and $\mathcal{T}$ is the linkage threshold for the whole dataset. 
Exploiting the linkage threshold, we propose Algorithm \ref{alg:Transform} to construct network layers with different linkage thresholds.

{\centering
\begin{minipage}{1.00\linewidth}
\begin{algorithm}[H]
\small
\caption{Data-to-Network Layers}\label{alg:Transform}
 \hspace*{\algorithmicindent} \textbf{Input:} $D$, a dataset of research collaboration.\\
 \hspace*{\algorithmicindent} \textbf{Output:} $\mathcal{G}$, a vector of generated network layers.\\
 
\begin{algorithmic}[1]
\Procedure{Transform-to-Network}{$D$}
\State $nodesList \gets \mathcal{ID}$ 
\State $\mathcal{P} \gets$ extract features from $D$
\State $trange \gets $ generate a space vector from $(\min(\mathcal{T})$, $\max(\mathcal{T}))$

    \For{each $threshold$ in $trange$}
        \For{each pair of $nodes$ in $nodesList$}
            \State $f \gets Aggregate(\mathcal{P}[nodes])$
            \If{$f \ge threshold$}
                \State $edgesList \gets$ pair of $nodes$
            \EndIf
            \State $Network\ G \gets GenNet (nodesList,edgesList)$
            \State Insert $G$ to $\mathcal{G}$
        \EndFor
    \EndFor
    \State return $\mathcal{G}$
\EndProcedure
\end{algorithmic}
\end{algorithm}
\end{minipage}
\par
}
\vspace{0.5cm}

\paragraph{Description of the algorithm:}
Assume $\mathcal{D}$ is the relational dataset of collaborations. We need to define the nodes and the links extracting particular features from $\mathcal{D}$ to generate the network layers. To define the list of nodes $nodesList$, we extract the $id$ of the entities which in this study we used the identities of research projects (see line $2$ in Algorithm \ref{alg:Transform}). In order to describe the links, we first need to define a linkage threshold $t$ considering the list of features $\mathcal{P}$. The linkage threshold of linear space vector consists of $n$ points of $threshold$ within the range of minimum and maximum values of $\mathcal{P}$ (see line $3-4$ in Algorithm \ref{alg:Transform}). Additionally, each network from $\mathcal{G}$ is measured by the set of network metrics. 

\begin{figure*}[h]
    \centering
    \includegraphics[width=0.9\textwidth]{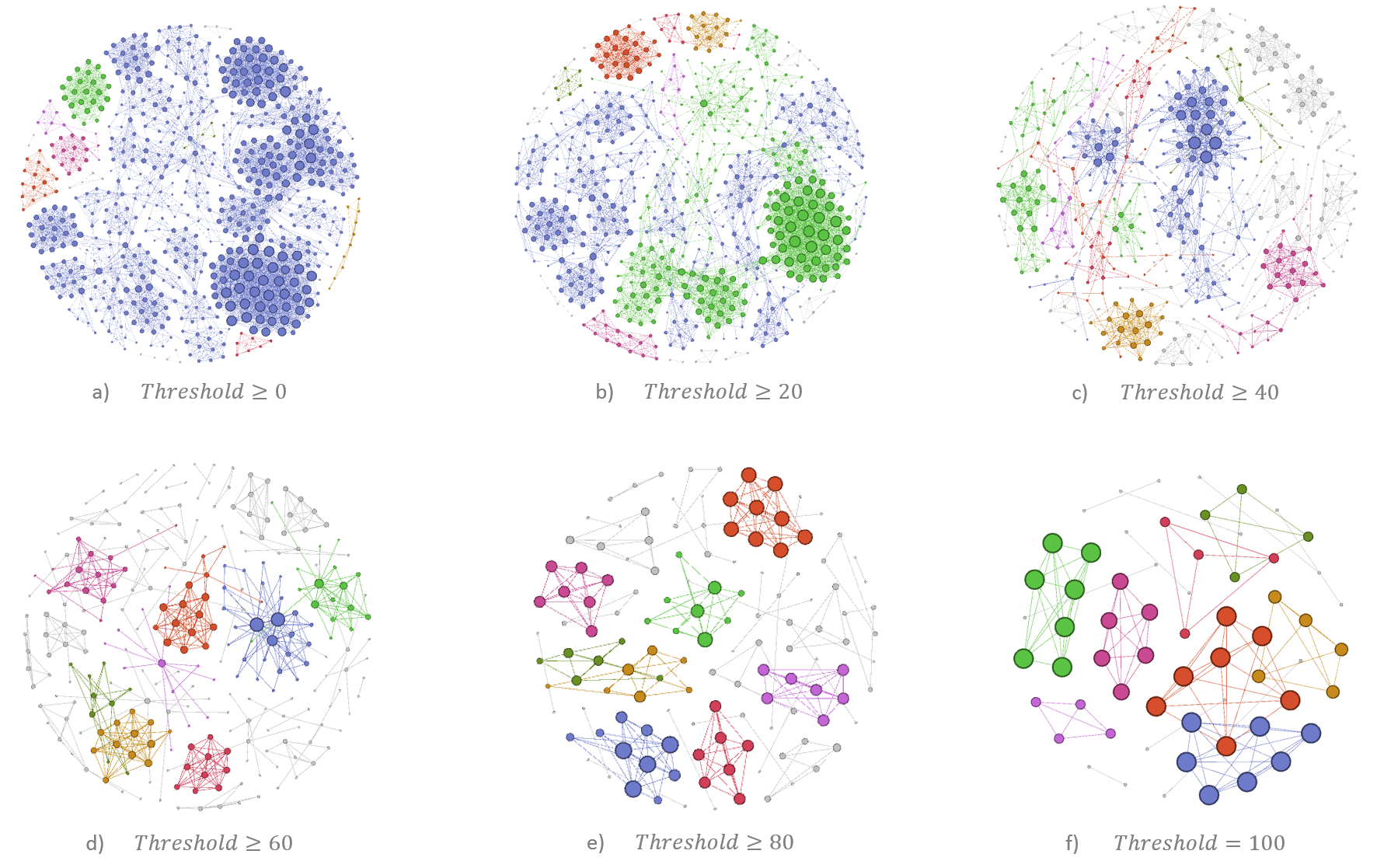}
    \caption{Visualization of the generated networks. In each network, the size of the nodes represents the degree of a node and the color illustrates the components. Such that blue shows components with the highest number of nodes, whereas gray represents the smallest components of a network. Moreover, green and red describe components which have a number of nodes within the range of previous cases.} 
    \label{fig:graphs}
\end{figure*}


We use the before mentioned collaboration data as input for the presented Algorithm \ref{alg:Transform}. Then, we measure the network metrics and utilize them for a better visualization of data. The resulting generated network layers are visualized in Fig. \ref{fig:graphs}. In this specific example, we consider $6$ thresholds for defining the edges starting from $0$ (the minimum collaboration observed in the dataset) to $100$ by using the connected components to enhance the visualization of data.
\paragraph{Complexity analysis} The complexity of Algorithm \ref{alg:Transform} depends on the two main parts of the algorithm: (1) Comparing each pair of nodes to find those that serve the determined condition for threshold is the most expensive one with the complexity of $O(n^2)$. (2) The complexity of the network generation is linear such that for $n$ nodes and $m$ edges the complexity is $O(n+m)$.


\section{Experiment Setup}\label{sec:setup}

\subsection{Dataset}

We benefit from a particular collaboration data derived from the \textit{National Electronics and Computer Technology Center (NECTEC)} that presents different projects and collaborations in the area of R\&D. The dataset is stored in a relational database consisting of research projects conducted between July 2013 and July 2018. Each project may consist of different deliverables: \textit{intellectual property (IP)}, \textit{papers}, or \textit{prototypes} as well as may comprise different members from different teams of the same institute. The dataset is the knowledge management about the project where the key information is to know (1) the type of the project, (2) project contributors and contributions.

The dataset of combined team tables have almost $8$k records which is the information of more than $2.3$k projects. Among them $630$ are related to IP, $1717$ to papers, and $539$ to prototypes. Overall, the institute has more than $1000$ members who are contributing on different projects with certain features (e.g., contribution percentage) which have been defined within the organization to evaluate the contributions. One of the main features we have used is the contribution percentage. The total percentage assigned to each project is $100\%$ that is divided between the project members according to their contribution on the project. Furthermore, IC-score is another feature that is developed by the institute and it illustrates the value of each project (e.g., prototype) based on its status (e.g., lab, industrial). To obtain the IC-score for each member, the total IC-score value of each project is divided by the contribution percentage of each member. The details regarding the values of both features (contribution percentage and IC-score) have been further discussed further in Section \ref{sec:results}.

\subsection{Network Metrics}

A network (or graph) $G = (V, E)$ consists of a set of vertices $V$ which are connected by the edges from set $E$. There exists different parameters (e.g., centrality measures) to analyze and study the networks. We choose centrality measures to analyze the generated network layers, which help to find the most important vertices within a network. Besides the centrality measures, we also consider other metrics such as network density and connected components to analyze the properties of the network. The following is a brief description of each metric.

\paragraph{Closeness Centrality} defines the closeness of a node to other nodes by measuring the average shortest path from that node to the all other vertices within the network. Hence, the more central a node is, the closer it is to all other nodes \cite{sabidussi1966centrality} calculated as \(C_C(v) = \sum_y{\frac{1}{d(v,u)}},\) where $d(v,u)$ is the distance between vertices $v$ and $u$.

\paragraph{Betweenness Centrality} indicates the number of times a node acts as a bridge along the shortest path between two other nodes. For a given node the number of shortest paths that passes through the node implies the betweenness centrality of the node. Nodes with high betweenness may have significant influence in a network due to their control over the flow of information passing between others through them. In a network $G = (V,E)$ betweenness centrality for node $v$ is \cite{brandes2001faster}:
        \( C_{B}(v) = \sum_{s\neq v \neq t \in V}{}\frac{\sigma_{st}(v)}{\sigma_{st}}\)
        where $\sigma_{st}$ total number of shortest paths from node $s$ to node $t$ and $\sigma_{st}(v)$ is the number of those paths that pass through $v$.
   
\paragraph{Degree Centrality} identifies the number of direct links which are connected to a vertex within the network. The importance of the nodes with higher degree is due to the immediate risk of these node while some information is flowing through the network. The degree of a node $v$ is represented as,
    \(C_{D}(v) = deg(v).\)
    
\paragraph{Clustering Coefficient} presents the likelihood of nodes in a network that tend to cluster together. The value of clustering coefficient lies between $0$ and $1$. When a network is clique which means that every two distinct vertices are adjacent, the value is $1$, however, in a star network in which a node's neighbours are not connected to each other at all, clustering coefficient is $0$. For an unweighted network, the clustering of a node $v$ is the fraction of possible triangles through that node that exist,
    \(CC(v) = \frac{2 T(v)}{deg(v)(deg(v)-1)}\)
    where $T(v)$ is the number of triangles through node $v$ and $deg(v)$ is the degree of $v$ \cite{saramaki2007generalizations}.
   
\paragraph{Network Density} is the ratio of potential links to existing links in a network. The range of this metric varies form $0$ for a network with no links (sparse network) and $1$ for networks with all possible links (dense network).
\(nd = \frac{2m}{n(n-1)},\)
where \(n\) is the number of nodes and \(m\) is the number of edges in network \(G\).
   
\paragraph{Connected Components} are sub-networks in which there are at least two vertices connected to each other through a path. In other words, two vertices are in the same sub-network if there is a path between them in the network. We use $n_{comp}$ as a notation to address the connected components in this the paper.

It is to be observed that the chosen metrics are (1) $local$ when they are only considering a node itself and the information of its neighbour to calculate, centrality measures are of this category, or (2) $global$ when they calculate a parameter considering the whole knowledge of network properties, nodes and edges, such as network density. 
Exploiting these metrics on networks generated from collaborative data, the purpose is to analyze the networks from different perspectives. Centrality measures are indicators to define the important node within a network. For instance, closeness centrality defines whether a project has a higher value for the institute such as delivering different outcomes (IP, papers, and prototypes). Moreover, there might be an argument regarding the topic of the project such that it is covering fundamental topics which other projects need to collaborate with. Degree centrality, represents the members' collaboration of a project with other projects. On the other hand, metrics like network density and connected components illustrates the general overview of a network. Network density represents how much a network is away from being a fully collaborative network in which all projects are connected together. 


\section{Results of the Network Analysis}\label{sec:results}

\subsection{Data Analysis}

\begin{figure*}
    \centering
    \includegraphics[width=0.9\textwidth]{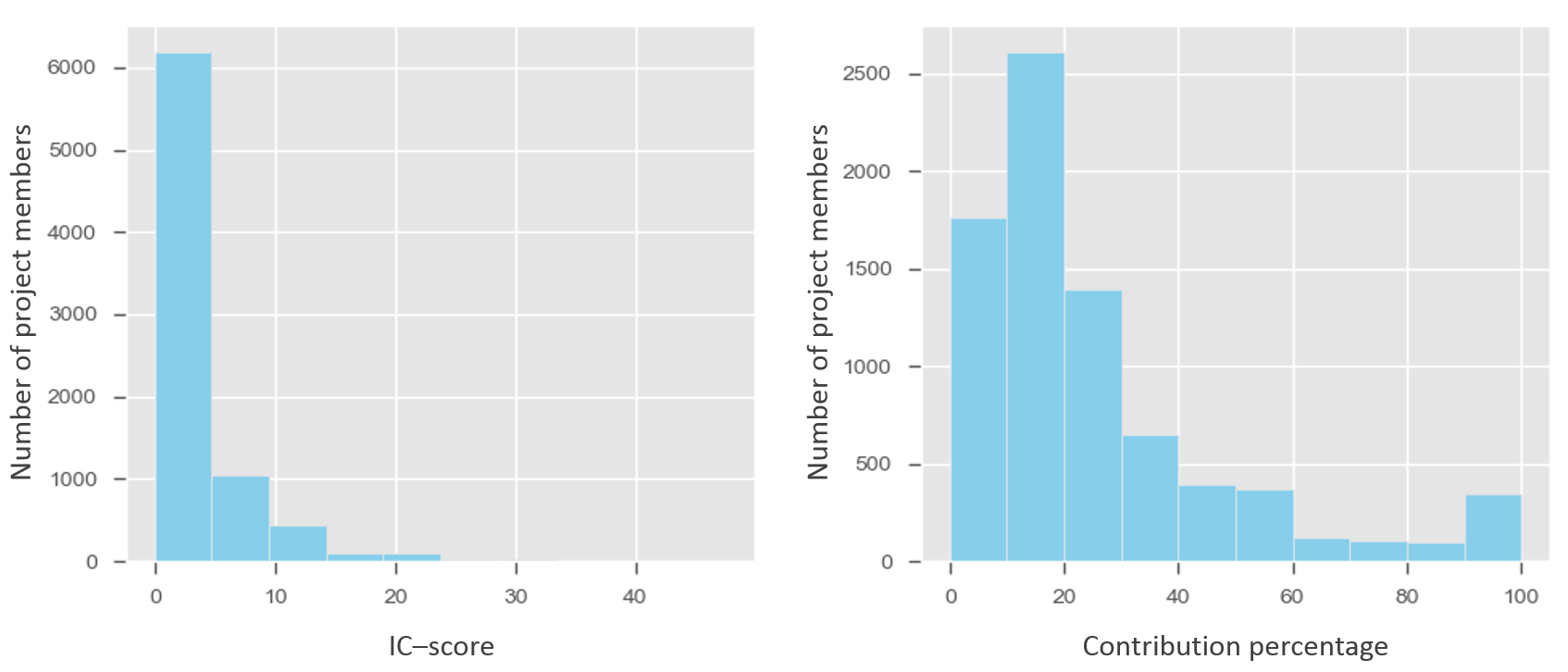}
   \caption{Histograms of IC-score and contribution percentage in the collaboration dataset.}
    \label{fig:histogram}
\end{figure*}

We perform a preliminary analysis on the data set in order to conduct on the linkage threshed. We exploit histograms to plot the frequency of the score and contribution percentage. Fig. \ref{fig:histogram} presents the histograms of IC-score and contribution percentage for each member in the collaboration data regarding all projects. The histograms represent the number of members with a certain value of IC-score (or contribution percentage) in the dataset. Moreover, mean, standard deviation, and variance are calculated for IC-score which are $3.16$, $4.24$, and $1.79$, respectively. For contribution percentage, the mean, standard deviation and variance are obtained as $23.30$, $22.80$ and $5.20$, respectively. IC-score concentrates on lower values better than contribution percentage, nevertheless, contribution percentage represent the dataset better as the coverage range is broader. Thus, we construct the linkage threshold defined in Equation (\ref{eq:linkagethreshold}) considering the contribution percentage. 

\subsection{Network Analysis}

We applied the proposed Algorithm \ref{alg:Transform} on our collaboration dataset. As a result, we obtained a vector of network layers, each represents a certain linkage threshold. For each network we calculated the set of network metrics which has been introduced in Section \ref{sec:setup}.

We chose $6$ linkage thresholds which are $0$, $20$, $40$, $60$, $80$, and $100$. For instance, with the linkage threshold equal to $20$ two projects in the network are connected if the average contribution percentage of the common members between those projects is equal or greater than $20\%$. Thus, those two projects are neighbours in the network. 
While increasing the threshold, the number of nodes that could not satisfy the condition increases dramatically. Thus, the number of isolated nodes increases which impact the outcomes of network metrics. In order to analyze the network regardless of the influence of these nodes, the network metrics are applied after removing the isolated nodes. We measured Betweeness Centrality $C_{B}(v)$, Degree Centrality $C_{D}(v)$, Closeness Centrality $C_{C}(v)$, Clustering Coefficient $CC(v)$, Network Density $nd$, and Connected Components $n_{comp}$ for each network. Additionally, for the \textit{local} metrics, we calculated the average of nodes for the whole network. 

\begin{figure*}
    \centering
    \includegraphics[width=\textwidth]{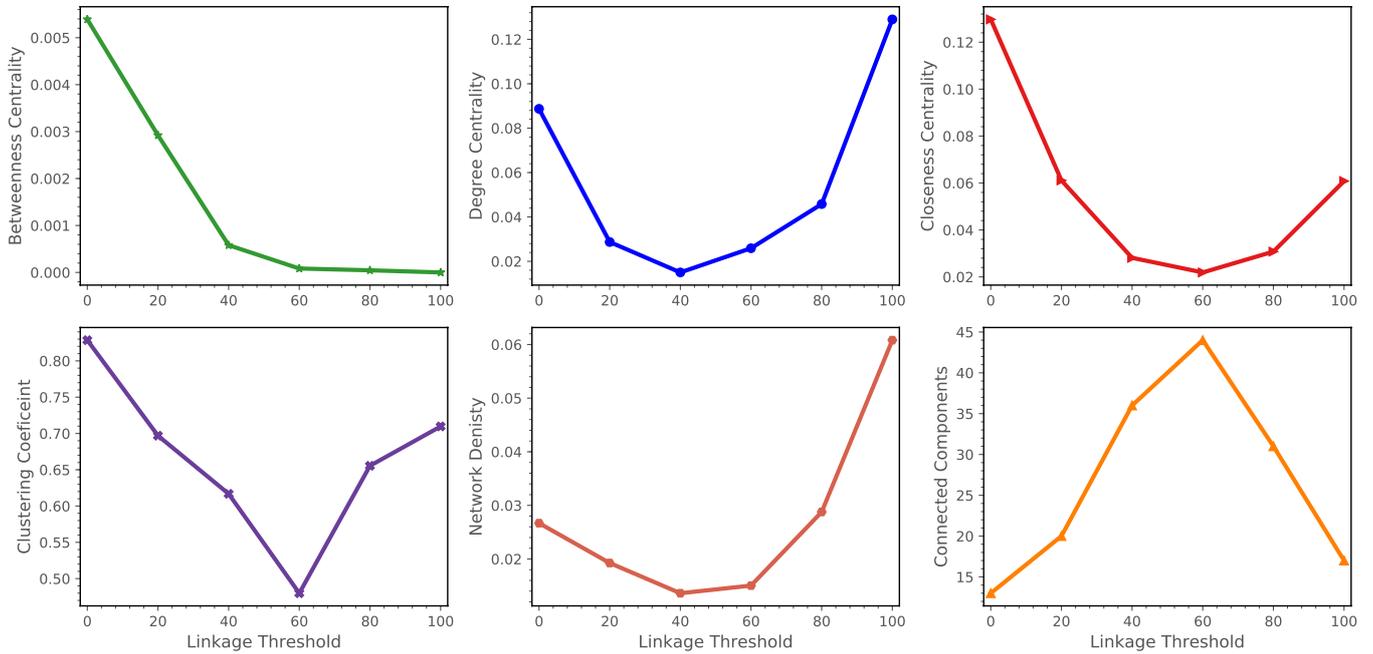}
    \caption{The effect of the set of network metrics, \({C_{C}(v), C_{B}(v), C_{D}(v), CC(v), nd, n_{comp}}\), on the generated network layers from IP projects. The result in each plot represents the average for all the nodes within a network layer. Thus, the x-axis illustrates the $linkage\ threshold$ in which the corresponding network has been created and the metric is measured.}
    \label{fig:metrics}
\end{figure*}

\begin{figure*}
    \centering
    \includegraphics[width=\textwidth]{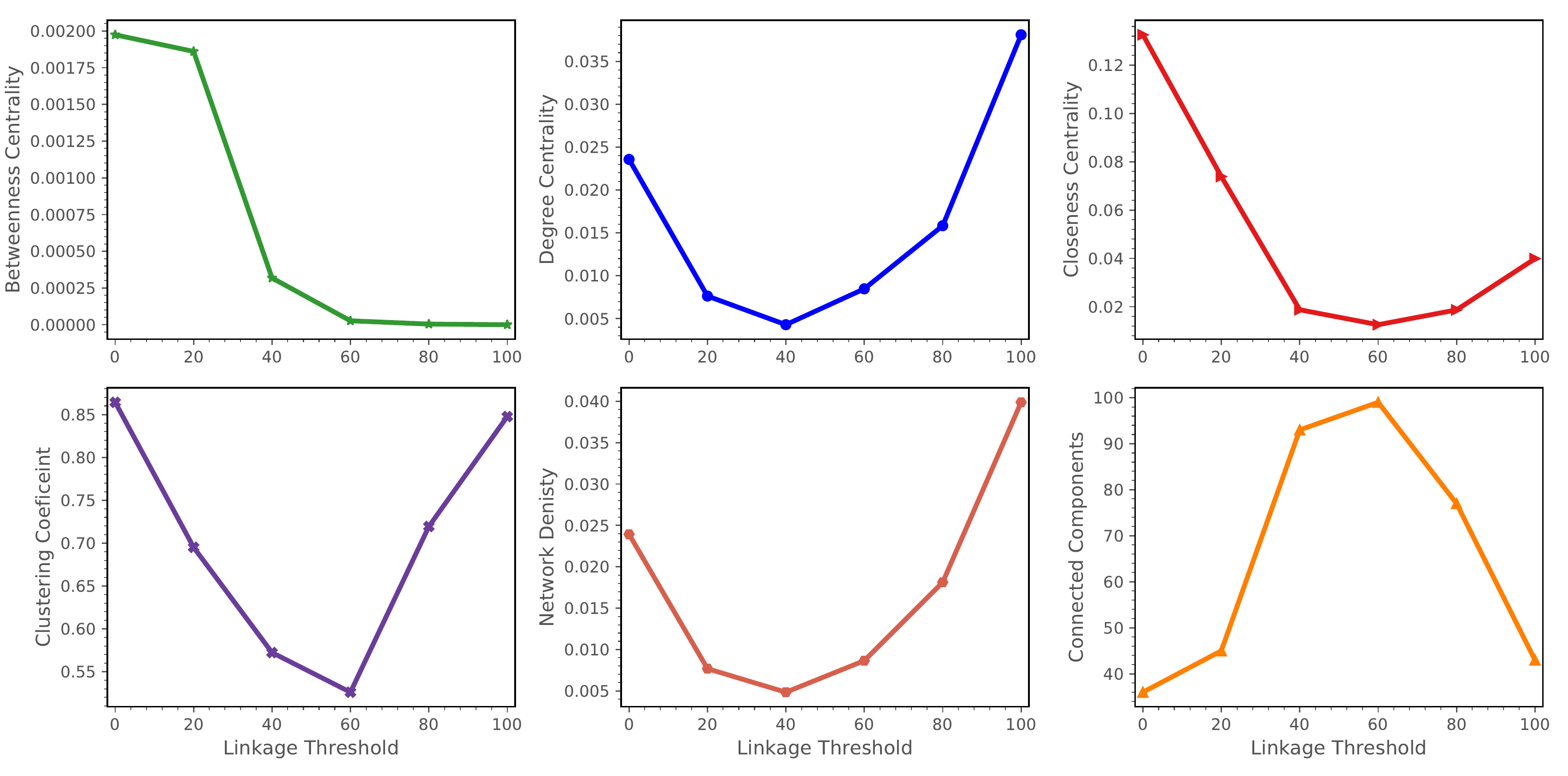}
    \caption{The effect of the set of network metrics on the generated network layers from IP, paper, and prototype projects.}
    \label{fig:metrics_all}
\end{figure*}

We first applied our algorithm considering only IP projects data, and then on the combination of all projects (i.e., IP, paper, and prototype). Fig. \ref{fig:metrics} describes the metrics on networks that are constructed particularly on IP dataset and Fig. \ref{fig:metrics_all} provides the results for a similar setup while considering the combination of all projects. 

According to Fig. \ref{fig:metrics} starting from $0$ as the linkage threshold increases the values of betweenness, degree and closeness centrality, clustering coefficient, network density decrease whereas the number of connected components increases until the linkage threshold is $60$. However, the pattern of the results for all metrics have dramatically changed within the range of $40$ to $60$. Besides the outcomes from Fig. \ref{fig:metrics}, that represents the result only for the IP projects, Fig. \ref{fig:metrics_all} provides the results of applying network metrics including all types of projects (IP, papers, and prototype). The results of both figures (Fig. \ref{fig:metrics} and \ref{fig:metrics_all}) are representing the similar patterns. In other words, the topology of networks are not any different from one project type to another. Furthermore, although the linkage threshold equal to $0$ provides detailed information of projects, the linkage threshold equal to $100$ represents a particular perspective of the dataset which describes the main leaders contribution in different projects. 

\subsection{Optimization and Future Work}

Our methodology to construct network layers from collaboration data reveals several optimization criteria. Optimizing the number of network layers while still containing the maximum on distinct information for enhanced analytics is a challenging task. Moreover, the linkage threshold we have defined in this paper can be generalized to a utility function to be performed on any given collaboration dataset. In addition, deciding on an optimal linkage threshold based on predefined criteria and conditions could further improve the performance, but also widen the applicability, of our algorithm. Additionally, we will consider different network representations for the same data in future work. We also plan on using more real-world collaboration data from distinct sources to further generalize our approach.

\section{Conclusion}\label{sec:conclusion}

The approach outlined in this paper infers possible collaboration networks of researchers within projects of an organization. Our method uses a linkage threshold to automatically generate these network layers from the relational input data. We conducted a network analysis on the produced networks using metrics such as clustering coefficient, closeness and betweenness centrality, and illustrate their impact on the different network layers. We, then, utilize the results of the metrics as an important input to visualize the generated graph in each configuration. We conclude that the linkage threshold has a crucial impact on the network properties and must be chosen with caution. Additionally, the influence of the linkage threshold on the results of the metrics indicates that the network representation can be optimized.
\\
\\
{\textbf{Acknowledgement.}}
This work is partially funded by the joint research programme UL/SnT-ILNAS on Digital Trust for Smart-ICT.

\balance

\bibliographystyle{abbrv}

\end{document}